\documentclass[journal,letterpaper,twocolumn,9pt]{article}

\usepackage{hyperref}
\usepackage{amssymb,amsmath}
\usepackage[noadjust]{cite}
\usepackage{graphicx} 
\usepackage{multirow}
\usepackage{arydshln}
\usepackage{epstopdf}
\usepackage{epsfig}
\usepackage{cite}
\usepackage{mathtools}
\usepackage{textcomp}
\usepackage{subfigure}

\usepackage[latin1]{inputenc}
\usepackage{color}

\usepackage[left=1.5cm, right=1.5cm, top=2.4cm, bottom=2.5cm]{geometry} 

\newcommand{\lcur}{\mathnormal{l}}
\newcommand{\gammab}{\boldsymbol{\theta}}
\newcommand{\kb}{\boldsymbol{k}}
\newcommand{\tb}{\boldsymbol{t}}

\newcommand{\dkb}{\boldsymbol{\Delta k}}
\newcommand{\ellb}{\boldsymbol{\ell}}
\newcommand{\omegab}{\boldsymbol{\omega}}
\newcommand{\Gamb}{{\varrho_j}}
\newcommand{\Clong}{{\varrho_j^{(1)}}}
\newcommand{\Cshort}{{\varrho_j^{(0)}}}
\newcommand{\Sigb}{\boldsymbol{\Sigma}_j}

\newcommand{\EE}{{\mathbb E}}
\newcommand{\RR}{{\mathbb R}}

\newcommand{\N}{\mathcal{N}}
\newcommand{\argmax}{\operatornamewithlimits{argmax}}
\newcommand{\mPSD}{\phi_j(\omegab;\gammab)}
\newcommand{\Admit}{\mathcal{A}}
\newcommand{\pri}{\pi}
\newcommand{\arp}{\pi}
\newcommand{\proposal}{p}
\newcommand{\mhp}{r}
\newcommand{\mi}{\mathrm{i}} 

\newcommand{\myln}{\log}

\newcommand{\Lead}{\ellb}
\newcommand{\llvec}{\boldsymbol{\lcur}_j}
\newcommand{\llead}{\lcur}
\newcommand{\Lvec}{\mathcal{L}}

\usepackage{color}
\definecolor{r}{rgb}{1,0,0}
\definecolor{b}{rgb}{1,0,0}
\definecolor{y}{rgb}{1,0,0}
\definecolor{green}{rgb}{0,0.6,0.4}
\definecolor{jy}{rgb}{0.7,0.3,0.3}

\title{\bf
Bayesian Estimation of the Multifractality Parameter for Image Texture Using a Whittle Approximation
}

\author{S. Combrexelle, H. Wendt$^*$, {\it Member, IEEE}, N. Dobigeon, {\it Senior Member, IEEE},\\ J.-Y. Tourneret, {\it Senior Member, IEEE}, S. McLaughlin, {\it Fellow, IEEE}, P. Abry, {\it Fellow, IEEE}
\thanks{%
This work was supported by ANR BLANC 2011 AMATIS BS0101102 and ANR Project Hypanema ANR-12-BS03-003. S. Combrexelle was supported by the Direction G\'en\'erale de l'Armement (DGA).
S. McLaughlin acknowledges the support of EPSRC grant number EP/J015180/1.%
}
\thanks{%
S. Combrexelle, N. Dobigeon, J.-Y. Tourneret and H. Wendt are with IRIT Laboratory, INP-ENSEEIHT, University of Toulouse, CNRS, Toulouse, France (email: firstname.lastname@enseeiht..fr) }
\thanks{%
S. McLaughlin is with the School of Engineering and Physical Sciences, Heriot-Watt University, Edinburgh, UK (email:  
s.mclaughlin@hw.ac.uk)}
\thanks{%
P. Abry is with the Physics Dept., Ecole Normale Sup\'erieure de Lyon, CNRS, France (email: patrice.abry@ens-lyon.fr)}
}

\providecommand{\keywords}[1]{\textbf{\textit{Index terms---}} #1}

\date{}

\begin{document}


\maketitle

\begin{abstract}

\boldmath\bf
Texture characterization is a central element in many image processing applications. Multifractal analysis is 
a useful  signal and image processing tool, yet, the accurate estimation of multifractal parameters for image texture remains a challenge. This is due in the main to the fact that current estimation procedures consist of performing linear regressions across frequency scales of the two-dimensional (2D) dyadic wavelet transform, for which only a few such scales are computable for images. The strongly non-Gaussian nature of multifractal processes, combined with their complicated dependence structure, makes it difficult to develop suitable models for parameter estimation. Here, we propose a Bayesian procedure that addresses the difficulties in the estimation of the multifractality parameter. 
The originality of the procedure is threefold: 
The construction of a generic semi-parametric statistical model for the logarithm of wavelet leaders; the formulation of Bayesian estimators that are associated with this model and the set of parameter values admitted by multifractal theory;
the exploitation of a suitable Whittle approximation within the Bayesian model which enables the otherwise infeasible evaluation of the posterior distribution associated with the model.
Performance is assessed numerically for several 2D multifractal processes, for several image sizes and a large range of process parameters.
The procedure yields significant benefits over current benchmark estimators in terms of estimation performance and ability to discriminate between the two most commonly used classes of multifractal process models.
The gains in performance are particularly pronounced for small image sizes, notably enabling for the first time the analysis of image patches as small as $64\times64$ pixels.

\end{abstract}



\section{Introduction}

%
%

\subsection{Context and motivation}
Since the early origins of digital image processing, texture has been recognized as one of the central characteristic features in images. There is no common definition for texture, and different paradigms have been introduced in the literature \cite{haralick1979}. 
Several authors have proposed to model texture using random fractals, scale invariance or self-similarity \cite{keller1989texture,PesquetPopescu2002}. Indeed, it has been reported in the literature that scale invariant processes are relevant and effective models for textures associated with a large class of natural images, see, e.g.,
\cite{IDCChainais,Wendt2009icip,Xu2010}.\\
The concepts of scale invariance and self-similarity are deeply tied to the degree of pointwise singular behavior or  \emph{local regularity} of the image amplitudes \cite{MandelIntTurb,Riedi2003}. It has long been recognized that  multiscale and wavelet analyzes constitute ideal tools to study data regularity \cite{muzyetal94,abfrv02,Jaffard2004,Riedi2003,Wendt20091100}.
It is therefore not surprising that these tools play a central role not only for the study of image contours (edges), but also for texture characterization
\cite{MallatThird,chang1993texture,unser1995texture}. 
Yet, while contours are essentially isolated singularities, the texture models consist of densely interwoven sets of singularities of different regularity strength. 
\emph{Multifractal analysis} provides a mathematical framework for the study of such \emph{spatial fluctuations of local regularity} and texture characterization is therefore nowadays often conducted using this tool \cite{VehelTexture1992,Arneodo2003a}.\\
\noindent{\bf Multifractal analysis.}\quad 
The local regularity of an image $X$ is commonly measured using the so-called \emph{H\"older exponent} $h(\tb)$ \cite{Jaffard2004,Riedi2003}. 
Qualitatively, the smaller $h(\tb_0)$, the rougher $X$  is at spatial location $\tb_0$ and the larger $h(\tb_0)$, the smoother $X$ is at $\tb_0$.
The goal of multifractal analysis is the estimation of the \emph{multifractal spectrum} $D(h)$, which provides a \emph{global} description of the spatial fluctuations of $h(\tb)$. It is defined as the collection of the fractal dimensions of the sets of points for which the H\"older exponent takes the same value \cite{Jaffard2004,Riedi2003}, cf., Section \ref{sec:MFA1}.
Multifractal analysis has recently matured into a standard image processing tool and has been successfully used in a large number of applications including texture classification \cite{Xu2010,Wendt2009icip}, biomedical applications \cite{Benhamou2001,Kestener2004}, physics \cite{Ponson2006,Roux2000} and art investigation \cite{jones2006fractal,coddington2008multifractal,Abry2013a,Johnson2014}.\\
In most applications, the estimation of $D(h)$ cannot be based directly on its definition
\cite{Jaffard2004}. 
Instead, a so-called \emph{multifractal formalism} is constructed based on multiresolution coefficients $T_X(a,\kb)$, essentially capturing the content of the image $X$ around the discrete spatial location $\kb$ for a given frequency scale $a=2^j$. Examples are given by increments, wavelet coefficients and more recently wavelet leaders $\Lead(j,\kb)$ \cite{Jaffard2004} (defined in Section \ref{sec:formalism}), which yield the current benchmark multifractal formalism.
The multifractal formalism provides an expansion of the multifractal spectrum of the image $X$ in terms of the so-called \emph{log-cumulants} $c_p$, $p\geq1$ \cite{Wendt20091100,Jaffard2014}
\begin{multline}
\label{equ:Dhcp}
 D(h) = 2 + \frac{c_2}{2!}\left( \frac{ h-{c_1}}{c_2}\right)^2+\frac{-{c_3}}{3!}\left( \frac{ h-{c_1}}{{c_2}}\right)^3
 \\
 +\frac{-{c_4}+3{c_3}^2/{c_2}}{4!}\left( \frac{ h-{c_1}}{{c_2}}\right)^4+ \ldots
\end{multline}
when $c_2 < 0$, while $D(h)=\delta(h-c_1)$ when $c_2\equiv0$ ($c_2$ cannot be positive theoretically \cite{LogSim,Jaffard2004,Riedi2003}). 

\noindent{\bf\boldmath Estimation of $c_2$.}\quad 
The leading order coefficients $c_p$ provide a relevant summary of the multifractal properties of $X$ in applications where it would often not be convenient to handle an entire function $D(h)$ \cite{LogSim,Wendt2007d,Wendt20091100,Jaffard2014}.
The first log-cumulant $c_1$, for instance, is the mode of $D(h)$ and can be read as a measure for the ``average'' smoothness of $X$. More importantly, the coefficient $c_2$, referred to as the \emph{multifractality} or \emph{intermittency} parameter, is directly related to the width of $D(h)$ and captures the multifractal signature (i.e., the fluctuations of the local regularity) of the image $X$. Its primary importance stems from the fact that it enables the identification of the two major classes of multifractal stochastic processes: self-similar processes for which $c_2= 0$ and multifractal multiplicative cascade (MMC) based processes for which $c_2$ is strictly negative \cite{MFASS}. 
While the former class is tied deeply to additive constructions, the latter is based on multiplicative constructions and is hence linked to fundamentally different physical principles \cite{TurbKolm,MandelIntTurb,Riedi2003}. Moreover, the magnitude of $c_2$ quantifies the degree of multifractality of an image for the latter class. For an overview and details on scale invariant and multifractal processes, the reader is referred to, e.g., \cite{Riedi2003,FractalPoint} and references therein.

In the seminal contribution \cite{LogSim}, it has been shown that the log-cumulants $c_p$ are tied to the quantities $\Lead(j,\kb)$ through the key relation
$
\text{Cum}_p[\myln \Lead(j,\kb)]=c_p^0+c_p\:\myln2^j,
$
where $\text{Cum}_p[\cdot]$ is the $p$-th order cumulant.
In particular
\begin{equation}
\label{eq:varj}
C_2(j) \triangleq \text{Var }[\myln \Lead(j,\kb)] = c_2^0+c_2\myln2^j.
\end{equation}
Relation \eqref{eq:varj} leads to the definition of the current standard and benchmark estimator for the parameter $c_2$, based on linear regression of the sample variance, denoted by $\widehat{\mbox{Var}}$, of $\myln \Lead(j,\kb)$ over a range of scales $j\in[j_1,j_2]$
\begin{equation}
\label{equ:c2LF}
\hat c_2 = \frac{1}{\myln2}\sum_{j=j_1}^{j_2}w_j \widehat{\mbox{Var}}[\myln \Lead(j,\cdot)]
\end{equation}
where $w_j$ are suitably defined regression weights \cite{Wendt20091100,Wendt2007d}.

\noindent{\bf Limitations.}\quad 
The use of multifractal analysis remains restricted to images of relatively large size (of order $512^2$ pixels) because a sufficient number of scales $j$ must be available to perform the linear regression \eqref{equ:c2LF}. While a similar issue is encountered for the analysis of 1D signals, it is significantly more severe for images: indeed, modulo border effects of the wavelet transform, the number of available scales is proportional to the logarithm of the number of samples for 1D signals and to the logarithm of the \emph{square root} of the number of pixels for an image. For instance, for a 1D signal with $256\times 256=65536$ samples, $j_{2}=13$ or $14$ scales can be computed, while $j_{2}=4$ or $5$ for an image of $N\times N=256\times256$ pixels. In addition, the finest scale, $j=1$, should not be used in \eqref{equ:c2LF}, see, e.g., \cite{Veitch2000}.
The practical consequences for the multifractal analysis of images are severe: 
First, images of small size and thus \emph{image patches} cannot be analyzed in practice. Second, \eqref{equ:c2LF} yields modest performance for images when compared with 1D signals of equivalent sample size \cite{Wendt20091100}, making it difficult to discriminate between $c_2\equiv0$ and values $c_2<0$ that are encountered in applications (typically, $c_2$ lies between $-0.01$ and $-0.08$).
The goal of this work is to propose and validate a novel procedure for the estimation of $c_2$ for images that addresses these difficulties.

\subsection{Related works}

There are a limited number of reports in the literature that attempt to overcome the limitations of multifractal analysis for images described above. 
The \textit{generalized method of moments} has been proposed and studied in, e.g., \cite{GMM_MFALux2007,GMM_MFALux2008,GMM_MFABacry2008} and formulates parameter inference as the solution (in the least squares sense) of an over-determined system of equations that are derived from the moments of the data. The method depends strongly on fully parametric models and yields, to the best of our knowledge, only limited benefits in practical applications.

Although classical in parameter inference, maximum likelihood (ML) and Bayesian estimation methods have mostly been formulated for a few specific self-similar  and multifractal processes \cite{Beran1994,wo92}.
The main reason for this lies in the complex statistical properties of most of these processes, which exhibit marginal distributions that are strongly  non-Gaussian as well as intricate algebraically decaying dependence structures that remain poorly studied to date. The same remark is true for their wavelet coefficients and wavelet leaders, see, e.g.,  \cite{Ossiander2000,ImpactVanish}.

One exception is given by the fractional Brownian motion (in 1D) and fractional Brownian fields (in 2D) (fBm), that are jointly Gaussian self-similar (i.e., $c_2\equiv0$) processes with fully parametric covariance structure appropriate for ML and Bayesian estimation. 
Examples of ML and Bayesian estimators for 1D fBm formulated in the spectral or wavelet domains can be found in \cite{wo92,Beran1994,chan2006estimation,moulines2008wavelet}.
For images, an ML estimator has been proposed in \cite{fBm_Bayes} (note, however, that the estimation problem is reduced to a univariate formulation for the rows / columns of the image there).

As far as MMC processes are concerned, \cite{ML_MRW} proposes an ML approach in the time domain for one specific process. However, the method relies strongly on the particular construction of this process and cannot easily accommodate more general model classes. Moreover, the method is formulated for 1D signals only.
Finally, a Bayesian estimation procedure for the parameter $c_2$ of multifractal time series has recently been proposed in \cite{Bayc2}. Unlike the methods mentioned above, it does not rely on specific assumptions but instead employs a heuristic semi-parametric model for the statistics of the logarithm of wavelet leaders associated with univariate MMC processes. Yet, it is designed for and can only be applied to univariate time series of small sample size. 

\subsection{Goals and contributions}

The use of fully parametric models for the data can be very restrictive in many real-world applications. 
Therefore, the goal and the main contribution of this work is to study a Bayesian estimation procedure for the multifractality parameter $c_2$ with as few as possible assumptions on the data (essentially, the relation \eqref{eq:varj}) that can actually be applied to real-world images of small as well as large sizes.
To this end, we adopt a strategy that is inspired by \cite{Bayc2} and develop the key elements that are required for its formulation for images.

First, we show by means of numerical simulations that the distribution of the logarithm of wavelet leaders $\myln \Lead(j,\kb)$ of 2D MMC processes can, at each scale $j$, be well approximated by a multivariate Gaussian distribution. Inspired by the covariance properties induced by the multiplicative nature of cascade constructions, we propose a new generic radial symmetric model for the variance-covariance of this distribution. This second-order statistical model is parametrized only by the two parameters $c_2$ and $c_2^0$ in \eqref{eq:varj} and enables us to formulate estimation in a Bayesian framework. 

Second, we formulate a Bayesian estimation procedure for the parameter $c_2$ of images that permits to take into account the constraints that are associated with the proposed statistical model. To this end, an appropriate prior distribution is assigned to the parameter vector $(c_2,c_2^0)$ which essentially ensures that the variance \eqref{eq:varj} is positive. Additional prior information, if available, can easily be incorporated.
The Bayesian estimators of $c_2$ associated with the posterior distribution of interest cannot be evaluated directly because of the constraints that the parameter vector $(c_2,c_2^0)$ has to satisfy. 
Therefore, we design a suitable Markov chain Monte Carlo (MCMC) algorithm that generates samples that are asymptotically distributed according to the posterior distribution of interest. These samples are in turn used to approximate the Bayesian estimators. More precisely, we propose a random-walk Metropolis-Hastings scheme to explore efficiently the posterior distribution according to the admissible set of values for $c_2$ and $c_2^0$.

Finally, the exact evaluation of the likelihood associated with the proposed model for the log-wavelet leaders requires the computation of the inverse and the determinant of large dense matrices, which is numerically and computationally too demanding for practical applications. 
To obtain a stable and efficient algorithm that can actually be applied to images, following intuitions developed in the univariate case, cf. e.g., \cite{Beran1994}, we approximate the exact likelihood with a Whittle-type expansion that is adapted to the proposed  model and can be efficiently evaluated in the spectral domain. 

The proposed algorithm for the estimation of the multifractality parameter $c_2$ is effective both for small and large image sizes. 
Its performance is assessed numerically by means of Monte Carlo simulations for two classical and representative 2D MMC constructions, the canonical Mandelbrot cascades (CMC) \cite{MandelIntTurb} and compound Poisson cascades (CPC) \cite{MultiProducts}, using the most common multipliers, and a large range of process parameters and sample sizes from $64\times64$ to $512\times512$ pixels.
Complementary results are provided for 2D fBms (that are self-similar but not MMC). Our results indicate that the proposed estimation procedure is robust with respect to different choices of process constructions and greatly outperforms \eqref{eq:varj}, in particular for small images and for identifying a value $c_2\equiv0$. It enables, for the first time, a multifractal analysis of images (or image patches) whose sizes are as small as $64\times64$ pixels.

The remainder of this work is organized as follows. 
Section \ref{sec:MFA} summarizes the main concepts of multifractal analysis and the wavelet leader multifractal formalism. Section \ref{sec:STAT} introduces the statistical model and the Bayesian framework underlying the estimation procedure for the parameter $c_2$ of images, which is formulated in Section \ref{sec:PRAC}. Numerical results are given in Section \ref{sec:RES}. In Section \ref{sec:Madonna}, the proposed procedure is applied to the patch-wise analysis of a real-world image, illustrating its potential benefits for practical applications. Finally, Section \ref{sec:conclusions} concludes this paper and presents some future work.

\section{Multifractal analysis of Images}
\label{sec:MFA}  
Let $X:\,\RR^2\to \RR$ denote the 2D function (image) to be analyzed.
The image $X$ is assumed to be locally bounded in what follows (see Section \ref{sec:formalism} for a practical solution to circumvent this prerequisite).
\subsection{Multifractal analysis}
\label{sec:MFA1} 

\noindent{\bf H\"older exponent.\quad} 
Multifractal analysis aims at characterizing the image $X$ in terms of the \emph{fluctuations} of its \emph{local regularity}, characterized by the so-called \emph{H\"older exponent}, which is defined as follows \cite{Jaffard2004,Riedi2003}.
The image $X$ is said to belong to $C^{\alpha}(\tb_0)$ if there exists $\alpha>0$ and a polynomial $P_{\tb_0}$ of degree smaller than $\alpha$ such that
\begin{align*}
\vert\!\vert X(\tb) - P_{\tb_0}(\tb)\vert\!\vert \le C\vert\!\vert \tb -\tb_0\vert\!\vert^{\alpha}
\end{align*}
where $\vert\!\vert\!\cdot\!\vert\!\vert$ is the Euclidian norm.
The {H\" older exponent} at position $\tb_0$ is the largest value of $\alpha$ such that this inequality holds, i.e.,
\begin{align}
h(\tb_0) \triangleq \sup\{\alpha: X \in C^{\alpha}(\tb_0)\}.
\end{align}

\noindent{\bf Multifractal spectrum.\quad} 
For large classes of stochastic processes, the H\"older exponents $h(\tb)$ can be theoretically shown to behave in an extremely erratic way \cite{Jaffard2004,Jaffard2014}. Therefore, 
multifractal analysis provides a \emph{global} description of the spatial fluctuations of $h(\tb)$ in terms of the \emph{multifractal spectrum} $D(h)$.
It is defined as the Hausdorff dimension (denoted $\dim_H$) of the sets of points at which the H\"older exponent takes the same value, i.e.,
\begin{equation}
\label{equ:Dh}
D(h)\triangleq\dim_H\big(E_h=\{\tb:h(\tb=h\}\big).
\end{equation}
For more details on multifractal analysis and a precise definition of the Hausdorff dimension, see, e.g., \cite{Jaffard2004,Jaffard2014}.

\subsection{Wavelet leader multifractal formalism}
\label{sec:formalism}

Historically, multifractal formalisms have been proposed based on increments or wavelet coefficients. These choices of multiresolution quantities lead to both theoretical and practical limitations, see \cite{Wendt2007d,Wendt20091100} for a discussion.
Recently, it has been shown that a relevant multifractal formalism can be constructed from  the \emph{wavelet leaders} \cite{Jaffard2004,Wendt2007d,Jaffard2014}, which are specifically tailored for this purpose.

\noindent\textbf{Wavelet coefficients.}\quad 
We assume that the image is given in form of discrete sample values $X(\kb)$, $\kb=(k_1,k_2)$.
A two-dimensional (2D) orthonormal discrete wavelet transform (DWT) can be obtained as the tensor product of one-dimensional (1D) DWT as follows. Let $G_0(k)$ and $G_1(k)$ denote the low-pass and high-pass filters defining a 1D DWT. These filters are associated with a mother wavelet $\psi$, characterized by its number of vanishing moments $N_{\psi}> 0$. Four 2D filters $G^{(m)}(\kb)$, $m=0,\dots,3$ are defined by tensor products of $G_i$, $i=1,2$. The 2D low-pass filter $G^{(0)}(\kb)\triangleq G_0(k_1)G_0(k_2)$ yields the approximation coefficients $D_X^{(0)}(j,\kb)$, whereas the high-pass filters defined by $G^{(1)}(\kb)\triangleq G_0(k_1)G_1(k_2)$, $G^{(2)}(\kb)\triangleq G_1(k_1)G_0(k_2)$ and $G^{(3)}(\kb)\triangleq G_1(k_1)G_1(k_2)$ yield the wavelet (detail) coefficients $D_X^{(m)}(j,\kb)$, $m=1,2,3$ as follows: at the finest scale $j=1$, the $D_X^{(m)}(j,\kb)$, $m=0,\dots,3$ are obtained by convolving the image $X$ with $G^{(m)}$, $m=0,\dots,3$, and decimation; for the coarser scales $j\geq 2$ they are obtained iteratively by convolving $G^{(m)}$, $m=0,\dots,3$, with $D_X^{(0)}(j-1,\cdot)$ and decimation.
For scaling and multifractal analysis purposes, the approximation  coefficients $D_X^{(0)}$ are discarded and it is common to normalize the wavelet coefficients according to the $L^1$-norm
\begin{align}
d_X^{(m)}(j,\kb)\triangleq2^{-j}D_X^{(m)}(j,\kb),\quad m=1,2,3
\end{align}
so that they reproduce the self-similarity exponent for self-similar processes \cite{Arneodo2003a}.
For a formal definition and details on (2D) wavelet transforms, the reader is referred to \cite{MallatThird,Antoine2004}.

\noindent\textbf{Wavelet leaders.}\quad 
Denote as
\begin{align*}
\lambda_{j,\kb}=\{[k_12^j,(k_1+1)2^j),[k_22^j,(k_2+1)2^j)\}
\end{align*}
the dyadic cube of side length $2^j$ centered at $\kb 2^j$ and
\begin{align*}
3\lambda_{j,\kb}=\bigcup_{n_1,n_2\in\{-1,0,1\}} \lambda_{j,k_1+n_1,k_2+n_2}
\end{align*}
the union of this cube with its eight neighbors. The wavelet leaders are defined as the largest wavelet coefficient magnitude within this neighborhood over all finer scales \cite{Jaffard2004}
\begin{align}
\label{equ:lx}
\Lead(j,\kb)\equiv \Lead(\lambda_{j,\kb}) \triangleq \sup_{m\in(1,2,3),\lambda'\subset3\lambda_{j,\kb}} \vert d_X^{(m)}(\lambda')\vert.
\end{align}
Wavelet leaders reproduce the H\"older exponent as follows
\begin{equation}
\label{equ:lxh}
h(\tb_0)=\liminf_{j\to-\infty}\big(\myln \Lead(\lambda_{j,\kb}(\tb_0))\big/\myln 2^j\big)
\end{equation}
where $\lambda_{j,\kb}(\tb_0)$ denotes the cube at scale $j$ including the spatial location $\tb_0$ \cite{Jaffard2004}. It has been shown that \eqref{equ:lxh} is the theoretical key property required for constructing a multifractal formalism, see \cite{Jaffard2004} for details.
In particular, it can be shown that the \emph{wavelet leader multifractal formalism} (WLMF), i.e., the use of \eqref{equ:Dhcp} with coefficients $c_p$ estimated using wavelet leaders, is valid for large classes of multifractal model processes, see \cite{Wendt2007d,Wendt20091100} for details and discussions.
The WLMF has been extensively studied both theoretically and in terms of estimation performance and constitutes the benchmark tool for performing multifractal analysis, cf. e.g., \cite{Wendt2007d,Wendt20091100}.

\noindent\textbf{Negative regularity.} \quad
The WLMF can be applied to locally bounded images (equivalently, to images with strictly positive uniform regularity) only, see  \cite{Wendt2007d,Wendt20091100,Abel2015} for precise definitions and for procedures for assessing this condition in practice. However, it has been reported that a large number of real-world images do not satisfy this prerequisite \cite{Wendt2009icip,Wendt20091100}. In these cases, a practical solution consists of constructing the WLMF using the modified wavelet coefficients 
\begin{equation}
\label{equ:fi}
d_X^{(m),\alpha}(j,\kb)\triangleq2^{\alpha j}d_X^{(m)}(j,\kb),\quad\alpha>0
\end{equation}
instead of $d_X^{(m)}$ in \eqref{equ:lx}. 
When $\alpha$ is chosen sufficiently large, the WLMF holds (see \cite{Wendt20091100} for details about the theoretical and practical consequences implied by this modification).

Finally, note that the above analysis as well as the WLMF are meaningful for \emph{homogeneous multifractal} functions $X$, for which the multifractal spectra $D(h)$ of different subsets of $\tb$ are identical. This excludes the class of \emph{multifractional} models \cite{Ayache2000covariance,Atto2014}, for which the function $h(\tb)$ is given by a smooth non-stationary evolution. 
Such models, also of interest in other application contexts, are not considered here, as the focus is on multifractality parameter $c_2$ which is not relevant to characterize multifractional processes.

\section{Bayesian framework}
\label{sec:STAT}

In this section, a novel empirical second-order statistical model for the logarithm of wavelet leaders for 2D MMC processes is proposed. This model is the key tool for estimating the multifractality parameter $c_2$ in a Bayesian framework.

\subsection{Modeling the statistics of log-wavelet leaders}
\begin{figure}[!t]
\centering
\includegraphics[width=1\linewidth]{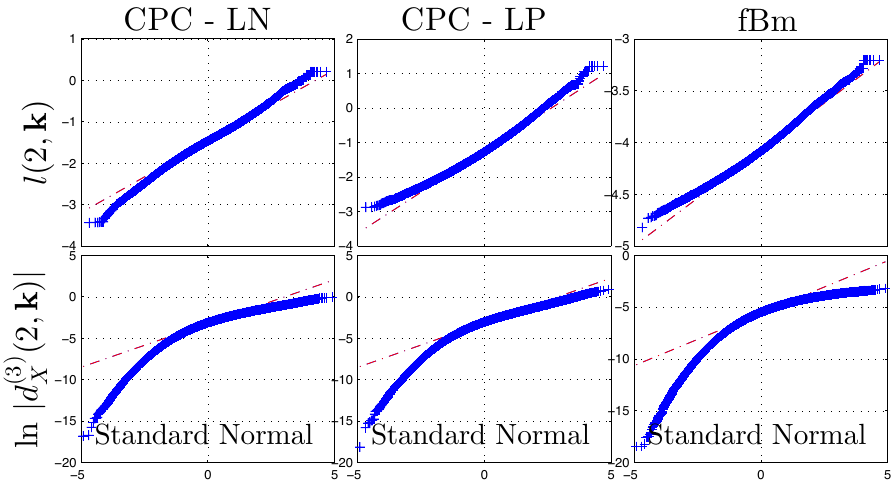}
\vskip-2mm
\caption{
Quantile-quantile plots of the empirical distributions of the log-wavelet leaders $\llead(2,\kb)$ (top) and wavelet coefficients $\myln d_X^{(3)}(2,\kb)$ (bottom)
against standard normal for CPC-LN (left column) and CPC-LP (center column) with $c_2=-0.04$, respectively,  and for fBm (right column).
}
\label{fig:qqplot}
\end{figure}

\noindent\textbf{Marginal distribution model.}\quad
It has recently been observed that for 1D signals the distribution of the \emph{log-wavelet leaders}
\begin{equation}
\llead(j,\kb)\triangleq\myln \Lead(j,\kb)
\end{equation}
can be reasonably well approximated by a Gaussian distribution \cite{Bayc2}. Here, we numerically investigate the marginal distributions of  $\llead(j,\cdot)$ for 2D images. 
To this end, a representative selection of scaling processes (the MMC processes CMC-LN, CMC-LP, CPC-LN and CPC-LP, as well as fBm, where LN stands for log-Normal and LP for log-Poisson, respectively) have been analyzed for a wide range of process parameters (see Section \ref{subsec:synthetic} for a description of these processes).
Representative examples of quantile-quantile plots of the standard Normal distribution against empirical distributions of log-wavelet leaders (scale $j=2$) associated with CPC-LN, CPC-LP and fBm are plotted in Fig. \ref{fig:qqplot} (upper row). 

Clearly, the normal distribution provides, within $\pm 3$ standard deviations, a reasonable approximation for the marginal distribution of log-wavelet leaders of images for both members of the MMC class.
It is also the case for the fBm, a Gaussian self-similar process that is not a member of MMC. 
Note that the fact that the marginal distributions of the log-wavelet leaders are approximately Gaussian for scale invariant processes confirms the intuitions formulated by Mandelbrot \cite{Mandelbrot1990}. However, it is not a trivial finding: There is no a priori reason for this property even if the analyzed stochastic process has log-normal marginals (as is the case for CMC-LN, for instance). Indeed, it is \emph{not} the case for the logarithm of the absolute value of wavelet coefficients whose marginal distributions are significantly more complicated and strongly depart from Gaussian, cf., Fig. \ref{fig:qqplot} (bottom row).

\begin{figure}[!t]
\centering
\includegraphics[width=1\linewidth]{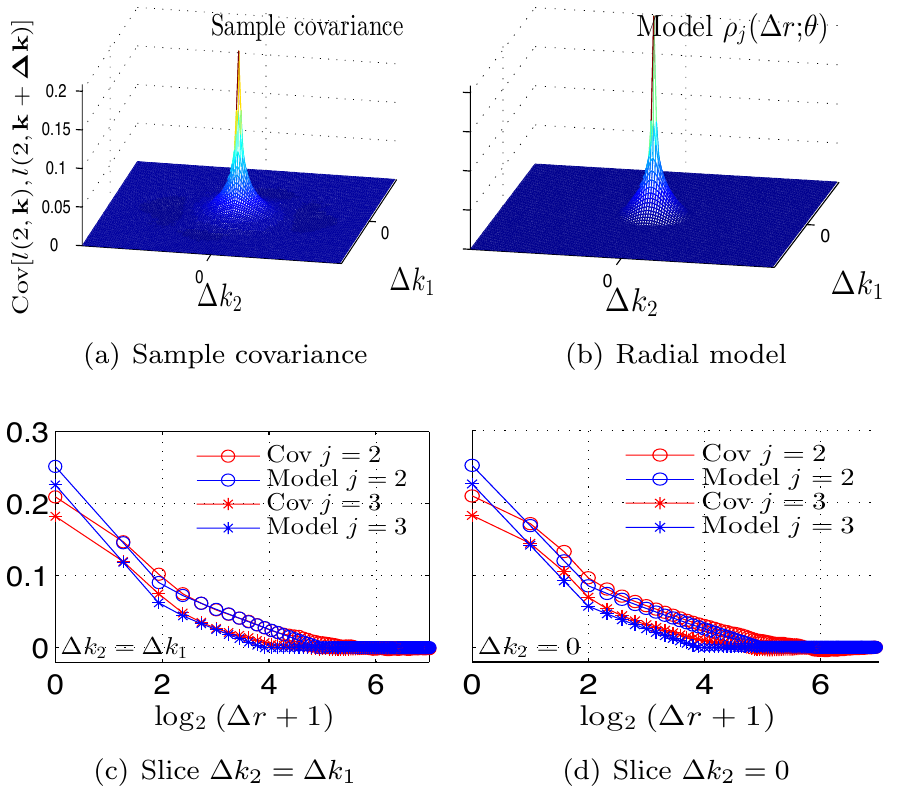}
\vskip-1mm
\caption{Fitting between the sample covariance (a), averaged on 100 realizations of CMC-LN ($[N,c_2]=[2^9,-0.04]$), and the parametric covariance (b);  (c) and (d) compare the model (blue) and the sample covariance (red) for two slices.}
\label{fig:cov}
\end{figure}

\noindent\textbf{Variance-covariance model.}\quad
We introduce a model for the covariance of the logarithm of 2D wavelet leaders for MMC processes at fixed scale $j$ denotated as $\text{Cov}[\llead(j,\kb),\llead(j,\kb+\dkb)]$. It is motivated by the asymptotic covariance of the logarithm of multiscale quantities generically associated with multiplicative construction (c.f. \cite{MandelIntTurb}), studied in detail for wavelet coefficients of 1D random wavelet cascades in \cite{Arneodo1998a}, and also by recent numerical results obtained for the covariance of the logarithm of 1D wavelet leaders for MMC processes \cite{Bayc2}. These results suggest a linear decay of $\text{Cov}[\llead(j,k),\llead(j,k+\Delta k)]$ in log coordinates $\myln \Delta k$, with slope given by the parameter $c_2$.
Numerical simulations with 2D MMC processes for a wide range of process parameters (detailed in Section \ref{subsec:synthetic}) indicate that the empirical intra-scale covariance is radially symmetric and decays as $c_2\myln\Delta r$ with $\Delta r\triangleq\vert\!\vert\dkb\vert\!\vert$ for an intermediary range of values $\Delta r$ given by $3< \Delta r \le \Delta r_j^\textrm{max}$ 
\begin{multline}
\label{equ:long}
\text{Cov}[\llead(j,\kb),\llead(j,\kb+\dkb)] \approx
\\\Clong(\Delta r;c_2) 
\triangleq \gamma +c_2(\log_2\:\Delta r+j)\myln 2
\end{multline} 
where $\Delta r_j^\textrm{max}=\sqrt{2}(\sqrt{n_j}-1)$ and $n_j \approx \lfloor N^2/2^{2j}\rfloor$ denotes the number of wavelet leaders at scale $2^j$ of an $N\times N$ image. The constant $\gamma$ is found to be well approximated by using the heuristic condition $\Clong(\lfloor \sqrt{n_j}/4\rfloor;c_2))=0$, where the operator $\lfloor\text{ }\rfloor$ truncates to integer values.

The theoretical variance of the log-wavelet leaders is given by $C_2(j)=C_2(j;c_2,c_2^0)$ defined in \eqref{eq:varj}.
Finally, the short-term covariance is modeled as a line connecting $C_2(j;c_2,c_2^0)$ at $\Delta r=0$ and $\Clong(\Delta r;c_2)$ at $\Delta r=3$ as follows
\begin{multline}
\label{equ:short}
\Cshort(\Delta r;c_2,c_2^0) \triangleq\\
\!\!\frac{\myln(\Delta r+1)}{ \myln4}[\Clong(3;c_2)-C_2(j;c_2,c_2^0)]+C_2(j;c_2,c_2^0).\!\!
\end{multline}
Combining \eqref{eq:varj}, \eqref{equ:long} and \eqref{equ:short} yields the following full model for the covariance, parametrized by two parameters $\gammab=[c_2,c_2^0]^T$ only
\begin{equation}
\Gamb(\Delta r;\gammab) = 
\begin{cases}
C_2(j;c_2,c_2^0) & \Delta r=0 \\
\Cshort(\Delta r;c_2,c_2^0) & 0 \le \Delta r \le 3 \\
\max(0,\Clong(\Delta r;c_2))\!\! & 3 \le \Delta r \le\Delta r_j^\textrm{max}.\!\!\!\!
\end{cases}
\label{equ:CovModel}
\end{equation}
Here, only the positive portions of $\Clong$ are considered for numerical reasons (conditioning of the covariance matrix).
The proposed covariance model is illustrated in Fig. \ref{fig:cov} for CMC-LN.

The joint Gaussian model with covariance model \eqref{equ:CovModel} assumes limited information on the dependence between different scales, essentially the variance \eqref{eq:varj}. The corresponding covariance matrix model for log-wavelet leaders at several scales $j\in[j_1,j_2]$ has thus block-diagonal structure. For convenience and without loss of generality, the formulations given below and in Section \ref{sec:PRAC} will be stated in block-diagonal form yet could be extended without difficulty to any other valid covariance matrix model.

\subsection{Likelihood, prior and posterior distributions}

We focus on the estimation of the parameter $c_2$ and therefore work with centered log-wavelet leaders below.
Let $\llvec$ denote the vector of the $n_j$ centered coefficients $\llead(j,\kb)-\widehat{\EE}[{\lcur}_X(j,.)]$ at scale  $j \in [j_1,j_2]$, organized in lexicographic order, where $\widehat{\EE}[\cdot]$ stands for the sample mean. 
Let $\Sigb(\gammab)$ denote the corresponding $n_j \times n_j$ covariance matrix whose entries are given by the 2D parametric covariance function model \eqref{equ:CovModel}.
For convenience of notation, all coefficients are stacked in a unique zero-mean  vector $\Lvec=[\boldsymbol{\lcur}_{j_1}^T,...,\boldsymbol{\lcur}_{j_2}^T]^T$.

\noindent\textbf{Likelihood.\quad}
With the above notation and assumptions, the likelihood of $\llvec$ is given by
\begin{equation}
\label{equ:LH}
p(\llvec|\gammab)\triangleq
\frac{
\exp\left(
-\frac{1}{2}\llvec^T\Sigb(\gammab)^{-1}\llvec
\right)
}{\sqrt{(2\pi)^{n_j}\det \Sigb(\gammab)}}.
\end{equation}
Using the independence between $\llead(j,\kb)$ and $\llead(j',\kb')$ for $j\neq j'$, the likelihood of $\Lvec$ is given by
\begin{equation}
p(\Lvec | \gammab) =\prod_{j=j_1}^{j_2} p(\llvec| \gammab).
\label{eq:exact}
\end{equation}

\noindent\textbf{Prior distribution.}\quad
The parameter vector $\gammab$ must be chosen such that the variances of  $\llead(j,\kb)$ are positive, $C_2(j)\ge0$.
We define the admissible set
\begin{equation}
\Admit=(\Admit^+\cup\:\Admit^-)\cap \Admit^m
\end{equation}
where 
$$
\begin{array}{rl}
\Admit^-&=\{(c_2,c_2^0)\in \mathbb{R}^2\mid c_2<0 \text{ and } c_2^0+c_2\:j_2\text{ln }2>0\},\\
\Admit^+&=\{(c_2,c_2^0)\in \mathbb{R}^2\mid c_2>0 \text{ and } c_2^0+c_2\:j_1\text{ln }2>0\},\\
\Admit^m&=\{(c_2,c_2^0)\in \mathbb{R}^2\mid |c_2^0|<c_2^{0,m}, |c_2|<c_2^{m} \}
\end{array}
$$
and $c_2^{m},c_2^{0,m}$ quantify the largest admissible values for $c_2$ and $c_2^0$, parameters that need to be tuned by practitioners and may depend on the application considered. 
When no additional prior information is available regarding $\gammab$, a uniform prior distribution on the set $\Admit$ is assigned to $\gammab$
\begin{equation}
\label{equ:prior}
\pri(\gammab) = U_{\Admit}(\gammab)\propto \mathbf{1}_{\Admit}(\gammab).
\end{equation}

\noindent\textbf{Posterior distribution and Bayesian estimators.}\quad The posterior distribution of $\gammab$ is obtained from the Bayes rule
\begin{align}
p(\gammab | \Lvec) \propto p(\Lvec | \gammab)\:\pri(\gammab)
\label{eq:post}
\end{align}
and can be used to define the Bayesian maximum a posteriori (MAP) and minimum mean squared error (MMSE) estimators given in \eqref{equ:map} and \eqref{equ:mmse} below.

\section{Estimation procedure}
\label{sec:PRAC}
The computation of the Bayesian estimators is not straight-forward because of the complicated dependence of the posterior distribution \eqref{eq:post} on the parameters $\gammab$. 
 Specifically, the inverse and determinant of $\Sigb$ in the expression of the likelihood \eqref{equ:LH} do not have a parametric form and hence \eqref{eq:post} can not be optimized with respect to the parameters $\gammab$.
In such situations, it is common to use a Markov Chain Monte Carlo (MCMC) algorithm generating samples that are distributed according to
$p(\gammab | \Lvec)$. These samples are used in turn to approximate the Bayesian estimators. 

\subsection{Gibbs sampler}
\label{subsec:Gibbs}
The following Gibbs sampler enables the generation of samples $\{\gammab^{(t)}\}_1^{N_{mc}}$ that are distributed according to the posterior distribution \eqref{eq:post}. This sampler consists of successively sampling according to the conditional distributions $p(c_2 | c_2^0, \Lvec)$ and $p(c_2^0 |c_2,\Lvec)$  associated with $p(\gammab|\Lvec)$.
To generate the samples according to the conditional distributions, a Metropolis-within-Gibbs procedure is used.
The instrumental distributions for the random walks are Gaussian and have variances $\sigma_{c_2}^2$ and  $\sigma_{c_2^0}^2$, respectively, which are adjusted to ensure an acceptance rate between $0.4$ and $0.6$ (to ensure good mixing properties).
For details on MCMC methods, the reader is referred to, e.g.,
\cite{MCMC_Robert}. 

\noindent\textbf{Sampling according to $p(c_2 | c_2^0,\:\Lvec)$.}\quad 
At iteration $t$, denote as $\gammab^{(t)}=[c_2^{(t)},c_2^{0,(t)}]^T$ the current state vector. 
A candidate $c_2^{(\star)}$ is drawn according to the proposal distribution $\proposal_1(c_2^{(\star)}| c_2^{(t)})= \N(c_2^{(t)},\sigma_{c_2}^2)$.
The candidate state vector $\gammab^{(\star)}=[c_2^{(\star)},c_2^{0,(t)}]^T$ is accepted with probability  $\arp_{c_2}=\min(1,\mhp_{c_2})$ (i.e., $\gammab^{(t+\frac{1}{2})}=\gammab^{(\star)}$) and rejected with probability $1-\arp_{c_2}$ (i.e., $\gammab^{(t+\frac{1}{2})}=\gammab^{(t)}$). Here, $\mhp_{c_2}$ is the Metropolis-Hastings acceptance ratio, given by
\begin{multline}
\label{equ:MHast}
\!\!\!\mhp_{c_2} \!=\! \frac{p(\gammab^{(\star)} | \Lvec)\:\proposal_1(c_2^{(t)}| c_2^{(\star)})}{p(\gammab^{(t)} | \Lvec)\:\proposal_1(c_2^{(\star)}| c_2^{(t)})} \!=\!
\mathbf{1}_{\Admit}(\gammab^{(\star)})\!\!
\prod_{j=j_1}^{j_2}
\sqrt{\frac{\det \Sigb(\gammab^{(t)})}{\det \Sigb(\gammab^{(\star)})}} \\
\times
\exp\left(
-\frac{1}{2}\llvec^T\left(\Sigb(\gammab^{(\star)})^{-1}-\Sigb(\gammab^{(t)})^{-1}\right)\llvec
\right).
\end{multline}

\noindent\textbf{Sampling according to $p(c_2^0 | c_2,\Lvec)$.}\quad Similarly, at iteration $t+\frac{1}{2}$, a candidate $c_2^{0,(\star)}$ is proposed according to the instrumental distribution $\proposal_2(c_2^{0,(\star)}| c_2^{0,(t)})= \N(c_2^{0,(t)},\sigma_{c_2^0}^2)$. The candidate state vector  $\gammab^{(\star)}=[c_2^{(t+\frac{1}{2})},c_2^{0,(\star)}]^T$ is accepted with probability $\arp_{c_2^0}=\min(1,\mhp_{c_2^0})$ (i.e., $\gammab^{(t+1)}=\gammab^{(\star)}$) and rejected with probability $1-\arp_{c_2^0}$ (i.e., $\gammab^{(t+1)}=\gammab^{(t+\frac{1}{2})}$). The Metropolis-Hastings acceptance ratio
$\mhp_{c_2^0}$ is given by \eqref{equ:MHast} with $p_1$ replaced by $p_2$, $c_2$ replaced by $c_2^0$ and $t$ replaced by $t+\frac{1}{2}$.

\noindent\textbf{Approximation of the Bayesian estimators.} \quad After a burn-in period defined by $t=1,\dots,N_{\textrm{bi}}$, the proposed Gibbs sampler generates samples  $\{\gammab^{(t)}\}_{t=N_{\textrm{bi}}+1}^{N_{\textrm{mc}}}$ that are distributed according to the posterior distribution (\ref{eq:post}). These samples are used to approximate the MAP and MMSE estimators
\begin{align}
\label{equ:map}
\hat \gammab^{\textrm{MMSE}} &\triangleq\EE[\gammab| \Lvec]\approx \frac{1}{N_{mc}-N_{bi}} \sum_{t=N_{bi}+1}^{N_{mc}} \gammab^{(t)} \\ 
\label{equ:mmse}
\hat \gammab^{\textrm{MAP}} &\triangleq\argmax_{\gammab} p(\gammab | \Lvec)\approx  \argmax_{t>N_{bi}} p(\gammab^{(t)}| \Lvec).
\end{align}

\subsection{Whittle approximation}
\label{subsec:Whittle}
The Gibbs sampler defined in subsection \ref{subsec:Gibbs} requires the inversion of the $n_j\times n_j$ matrices $\Sigb(\gammab)$ in \eqref{equ:MHast} for each sampling step in order to obtain $\mhp_{c_2}$ and $\mhp_{c_2^0}$. 
These inversion steps are computationally prohibitive even for very modest image sizes (for instance, a $64\times64$ image would require the inversion of a dense matrix of size $\sim 1000\times1000$ at scale $j=1$ at each sampling step). In addition, it is numerically instable for larger images (due to growing condition number).
To alleviate this difficulty, we propose to replace the exact likelihood \eqref{eq:exact} with an asymptotic approximation due to Whittle \cite{ApproxLikeIrr0,ApproxLikeIrr}.
With the above assumptions, the collection of log-leaders $\{\llead(j,\cdot)\}$ are realizations of a Gaussian random field on a regular lattice $P_j=\{1,..,m_j\}^2$, where $m_j=\sqrt{n_j}$. 
Up to an additive constant, the Whittle approximation for the negative logarithm of the Gaussian likelihood \eqref{equ:LH} reads \cite{ApproxLikeIrr,ApproxLikeIrr2,Beran1994,ParamEstRF}

\begin{multline}
-\myln p(\llvec|\gammab)\approx \\
p_W(\llvec|\gammab) =\frac{1}{2}\sum_{\omegab \in D_j} \myln\mPSD+\frac{I_j(\omegab)}{n_j\mPSD}
\label{eq:Whittle}
\end{multline}
where the summation is taken over the spectral grid $D_j= \{\frac{2\pi}{m_j}\lfloor(-m_j-1)/2 \rfloor,-1,1,m_j-\lfloor m_j/2\rfloor\}^2$.
Here $I_j(\omegab)$ is the 2D standard periodogram of $\{\llead(j,\kb)\}_{\kb\in P_j}$ 
 \begin{equation}
 I_j(\omegab)=\bigg\vert\sum_{\kb\in P_j}\llead(j,\kb)\exp(-\mi\kb^{T}\omegab)\bigg\vert^2
 \end{equation}
and $\mPSD$ is the spectral density associated with the covariance function $\Gamb(\Delta r;\gammab)$, respectively.  
Without a closed-form expression for $\mPSD$, it can be evaluated numerically by discrete Fourier transform (DFT)
\begin{equation}
\label{covpsd}
\mPSD = \bigg| \sum_{\kb\in P_j} \Gamb(\sqrt{k_1^2+k_2^2};\gammab) \exp(-\mi\kb^{T}\omegab) \bigg|.
\end{equation}

\begin{figure}[!t]
\centering
\includegraphics[width=1\linewidth]{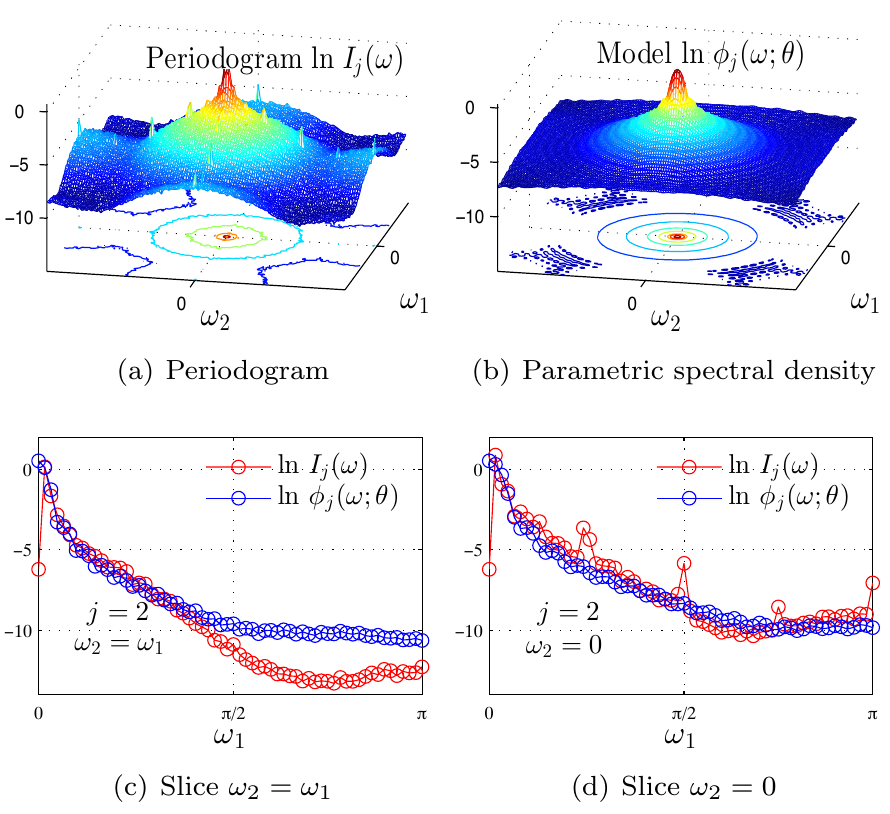}\vspace{-2mm}
\caption{Fitting between the periodogram (a), averaged on 100 realizations of CMC-LN ($[N,c_2,j]=[2^
9,-0.04,2]$), and the  model $\mPSD$, obtained from a DFT of $\Gamb(\Delta r;\gammab)$. (c) and (d) compare the model (blue) and the periodogram (red) for two slices.\label{fig:psd}}
\end{figure}

\noindent\textbf{Frequency range.} \quad
It is commonly reported in the literature that the range of frequencies used in \eqref{eq:Whittle}  can be restricted. This is notably the case for the periodogram-based estimation of the memory coefficient of long-range dependent time series for which only the low frequencies can be used, see, e.g., \cite{Beran1994,WhittlePseudoMax,GaussianLRD}.
Similarly, in the present context, the proposed spectral density model $\mPSD$ yields an excellent fit at low frequencies and degrades at higher frequencies. This is illustrated in Fig. \ref{fig:psd} where the average periodograms of $\llead(j,\kb)$ for CPC-LN
are plotted together with the model $\mPSD$. 
We therefore restrict the summation in \eqref{eq:Whittle} to low frequencies $D_j^{\dagger}(\eta)=\left\{\omegab\in D_j | \|\omegab\|_2\le \sqrt{\eta}\frac{2\pi}{m_j}\lfloor m_j/2\rfloor\right\}$ where the fixed parameter $\eta$ approximately corresponds to the fraction of the spectral grid $D_j$ that is actually used. 
We denote the approximation of $p_W(\llvec|\gammab)$ obtained by replacing $D_j$ with $D_j^{\dagger}$ in \eqref{eq:Whittle} by $p_W^\dagger(\llvec|\gammab)$.
The final Whittle approximation of the likelihood \eqref{eq:exact} is then given by the following equation
\begin{multline}
p(\Lvec | \gammab) \approx\exp\left(-\sum_{j=j_1}^{j_2}p_W^\dagger(\llvec|\gammab)\right)\\
\!\!\!=\exp\!\left(\!-\!\sum_{j=j_1}^{j_2} \frac{1}{2}\!\!\sum_{\omegab \in D_j^\dagger(\eta)}\!\!\! \myln\mPSD+\frac{I_j(\omegab)}{n_j\mPSD}\!\right)\!
\label{fullWhittleMod}
\end{multline}
up to a multiplicative constant.
It is important to note that the restriction to low frequencies in \eqref{fullWhittleMod} is not a restriction to a limited frequency content of the image (indeed, scales $j\in[j_1,j_2]$ are used) but only concerns the numerical evaluations of the likelihood \eqref{eq:exact}.

\begin{figure}[t]
\centering
\includegraphics[width=1\linewidth]{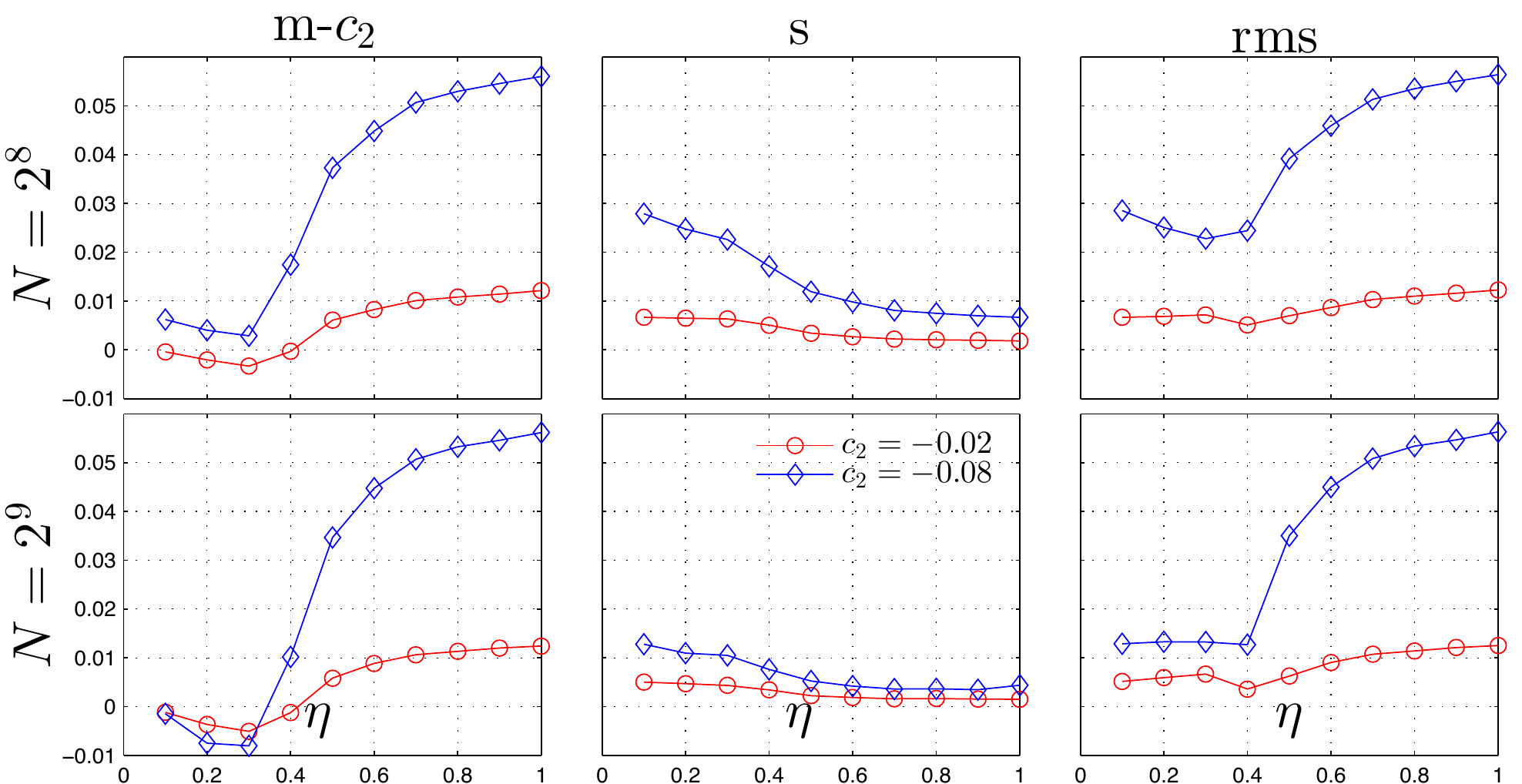}\vspace{-1mm}
\caption{Influence of the bandwidth parameter, $\eta$, on estimation performance for two different sizes of LN-CMC and two different values of $c_2$ values ($-0.02$ in red and $-0.08$ in blue).}
\label{fig:bandwidth}
\end{figure}

\section{Numerical experiments}
\label{sec:RES}
The proposed algorithm is numerically validated for several types of scale invariant and multifractal stochastic processes for different sample sizes and a large range of values for  $c_2$.

\subsection{Stochastic multifractal model processes}
\label{subsec:synthetic}
\noindent\textbf{Canonical Mandelbrot Cascade (CMC).}\quad CMCs \cite{MandelIntTurb} are the historical archetypes of multifractal measures. 
Their construction is based on an iterative split-and-multiply procedure on an interval; we use a 2D binary cascade for two different multipliers: 
First, log-normal multipliers $W=2^{-U}$, where $U\sim \mathcal{N}(m,2m/\myln2)$ is a Gaussian random variable (CMC-LN); Second, log-Poisson multipliers $W=2^\gamma\exp\left(\myln(\beta)\pi_\lambda\right)$, where $\pi_\lambda$ is a Poisson random variable with parameter $\lambda=-\frac{\gamma\myln2}{(\beta-1)}$ (CMC-LP).
For CMC-LN, 
the log-cumulants are given by $c_{1}=m+\alpha$, $c_2=-2m$ and $c_p= 0$ for all $p\geq3$. 
For CMC-LP,
$c_1=\alpha+\gamma\left(\frac{\myln(\beta)}{\beta-1}-1\right)$ and all higher-order log-cumulants are non-zero with
$c_p=-\frac{\gamma}{\beta-1}\left(-\myln(\beta)\right)^p$, $p\geq2$. Below, $\gamma = 1.05$ and $\beta$ is varied according to the value of $c_2$.

\begin{table}[h!]
\setlength{\tabcolsep}{0.7mm}
\renewcommand{\arraystretch}{0.95}
\centering
\caption{\label{tab:LARGE}Estimation performance for CMC-LN ($\textnormal{a}$), CPC-LN ($\textnormal{b}$) CMC-LP ($\textnormal{c}$), CPC-LP ($\textnormal{d}$) for sample sizes $N=\{2^8,2^9\}$ and $j_1=2$, $j_2=\{4,5\}$. Best results are marked in bold.}
\begin{tabular}{c|c|c||r|r|r|r|r}
    \hline
\multicolumn{8}{l}{(a)\hskip3.2cm CMC-LN}\\
\hline
\multicolumn{3}{c||}{$c_2$}&$ -0.01\;\;$&$ -0.02\;\;$&$ -0.04\;\;$&$ -0.06\;\;$&$ -0.08\;\;$\\ 
\hline
    \parbox[t]{2mm}{\multirow{6}{*}{\rotatebox[origin=c]{90}{ $N=2^8$}}}&\parbox[t]{1mm}{\multirow{2}{*}{\rotatebox[origin=c]{90}{m}}} & LF &$-0.015$&$-0.027$&$-0.053$&$-0.066$&$-0.087$\\
	   &  & MMSE &$\mathbf{-0.014}$&$\mathbf{-0.023}$&$\mathbf{-0.042}$&$\mathbf{-0.060}$&$\mathbf{-0.078}$\\
	        \cline{2-8}
	   & \parbox[t]{1mm}{\multirow{2}{*}{\rotatebox[origin=c]{90}{s}}}& LF &$ 0.010$&$ 0.011$&$ 0.015$&$ 0.018$&$ 0.030$\\
	  &  & MMSE &$\mathbf{0.005}$&$\mathbf{0.006}$&$\mathbf{0.013}$&$\mathbf{0.014}$&$\mathbf{0.020}$\\
	        \cline{2-8}
	   & \parbox[t]{1mm}{\multirow{2}{*}{\rotatebox[origin=c]{90}{rms}}}& LF &$ 0.011$&$ 0.014$&$ 0.019$&$ 0.019$&$ 0.030$\\
	 &   & MMSE &$\mathbf{0.007}$&$\mathbf{0.007}$&$\mathbf{0.013}$&$\mathbf{0.014}$&$\mathbf{0.020}$\\
	    \hline
	     \hline
    \parbox[t]{2mm}{\multirow{6}{*}{\rotatebox[origin=c]{90}{ $N=2^9$}}}&\parbox[t]{1mm}{\multirow{2}{*}{\rotatebox[origin=c]{90}{m}}}& LF &$-0.016$&$-0.027$&$-0.049$&$ -0.070$&$\mathbf{-0.087}$\\
	   & & MMSE &$\mathbf{-0.014}$&$\mathbf{-0.025}$&$\mathbf{-0.047}$&$\mathbf{-0.067}$&$\mathbf{-0.087}$\\
	        \cline{2-8}
	   & \parbox[t]{1mm}{\multirow{2}{*}{\rotatebox[origin=c]{90}{s}}}& LF &$ 0.005$&$ 0.006$&$ 0.008$&$ 0.011$&$ 0.016$\\
	   & & MMSE &$\mathbf{0.002}$&$\mathbf{0.004}$&$ \mathbf{0.006}$&$\mathbf{0.008}$&$\mathbf{0.012}$\\
	        \cline{2-8}
	   & \parbox[t]{1mm}{\multirow{2}{*}{\rotatebox[origin=c]{90}{rms}}}& LF &$ 0.008$&$ 0.010$&$ 0.012$&$ 0.015$&$ 0.018$\\
	   &  & MMSE &$\mathbf{0.005}$&$\mathbf{0.007}$&$ \mathbf{0.009}$&$\mathbf{0.011}$&$\mathbf{0.014}$\\
    \hline
 \hline 
\multicolumn{8}{l}{(b)\hskip3.2cm CPC-LN}\\
\hline
\multicolumn{3}{c||}{$c_2$}&$ -0.01\;\;$&$ -0.02\;\;$&$ -0.04\;\;$&$ -0.06\;\;$&$ -0.08\;\;$\\ 
\hline
\parbox[t]{2mm}{\multirow{6}{*}{\rotatebox[origin=c]{90}{ $N=2^8$}}}&\parbox[t]{1mm}{\multirow{2}{*}{\rotatebox[origin=c]{90}{m}}}& LF&$-0.013$&$\mathbf{-0.025}$&$-0.049$&$\mathbf{-0.066}$&$-0.089$\\ 
 & & MMSE&$\mathbf{-0.007}$&$-0.013$&$\mathbf{-0.035}$&$\mathbf{-0.054}$&$\mathbf{-0.074}$\\ 
\cline{2-8}
& \parbox[t]{1mm}{\multirow{2}{*}{\rotatebox[origin=c]{90}{s}}}& LF&$0.009$&$ 0.011$&$ 0.017$&$ 0.024$&$ 0.029$\\ 
 & & MMSE&$\mathbf{0.003}$&$\mathbf{0.005}$&$\mathbf{0.011}$&$ \mathbf{0.016}$&$\mathbf{0.022}$\\ 
\cline{2-8}
& \parbox[t]{1mm}{\multirow{2}{*}{\rotatebox[origin=c]{90}{rms}}}& LF&$0.010$&$ 0.012$&$  0.020$&$ 0.025$&$ 0.031$\\ 
 & & MMSE&$\mathbf{0.004}$&$\mathbf{0.008}$&$\mathbf{0.012}$&$ \mathbf{0.018}$&$\mathbf{0.023}$\\ 
\hline
\hline
\parbox[t]{2mm}{\multirow{6}{*}{\rotatebox[origin=c]{90}{ $N=2^9$}}}&\parbox[t]{1mm}{\multirow{2}{*}{\rotatebox[origin=c]{90}{m}}}& LF&$-0.013$&$-0.025$&$-0.045$&$-0.066$&$-0.089$\\ 
 & & MMSE&$\mathbf{-0.008}$&$\mathbf{-0.015}$&$\mathbf{-0.035}$&$\mathbf{-0.057}$&$\mathbf{-0.079}$\\ 
\cline{2-8}
& \parbox[t]{1mm}{\multirow{2}{*}{\rotatebox[origin=c]{90}{s}}}& LF&$0.004$&$0.005$&$0.010$&$ 0.014$&$ 0.015$\\ 
 & & MMSE&$\mathbf{0.002}$&$\mathbf{0.003}$&$\mathbf{0.006}$&$\mathbf{0.009}$&$ \mathbf{0.013}$\\ 
\cline{2-8}
& \parbox[t]{1mm}{\multirow{2}{*}{\rotatebox[origin=c]{90}{rms}}}& LF&$0.005$&$0.007$&$ 0.011$&$ 0.015$&$ 0.017$\\ 
 & & MMSE&$\mathbf{0.003}$&$\mathbf{0.005}$&$\mathbf{0.008}$&$ \mathbf{0.009}$&$\mathbf{0.013}$\\
 \hline 
   \end{tabular}
\vskip1mm
\setlength{\tabcolsep}{0.2mm}
\begin{tabular}{c|c|c||r|r|r||r|r|r}
\hline 
\multicolumn{6}{l||}{\hskip1.7cm (c)\hskip0.9cm CMC-LP}&\multicolumn{3}{l}{(d)\hskip0.9cm CPC-LP}\\
\hline
\multicolumn{3}{c||}{$c_2$}&$ -0.02\;\;$&$ -0.04\;\;$&$ -0.08\;\;$ &$ -0.02\;\;$&$ -0.04\;\;$&$ -0.08\;\;$\\ 
\hline
\parbox[t]{2mm}{\multirow{6}{*}{\rotatebox[origin=c]{90}{ $N=2^8$}}}&\parbox[t]{1mm}{\multirow{2}{*}{\rotatebox[origin=c]{90}{m}}}& LF&$\mathbf{-0.019}$&$\mathbf{-0.038}$&$\mathbf{-0.076}$&$-0.043$&$-0.065$&$ -0.120$\\ 
 & & MMSE&$-0.017$&$-0.032$&$-0.063$&$\mathbf{-0.029}$&$\mathbf{-0.055}$&$\mathbf{-0.100}$\\ 
\cline{2-9}
& \parbox[t]{1mm}{\multirow{2}{*}{\rotatebox[origin=c]{90}{s}}}& LF&$0.010$&$ 0.014$&$ 0.023$&$ 0.016$&$ 0.035$&$ 0.035$\\ 
 & & MMSE&$\mathbf{0.005}$&$\mathbf{0.009}$&$ \mathbf{0.016}$&$\mathbf{0.010}$&$\mathbf{0.012}$&$\mathbf{0.027}$\\ 
\cline{2-9}
& \parbox[t]{1mm}{\multirow{2}{*}{\rotatebox[origin=c]{90}{rms}}}& LF&$0.010$&$ 0.014$&$ 0.023$&$ 0.028$&$ 0.043$&$  0.050$\\ 
 & & MMSE&$\mathbf{0.006}$&$\mathbf{0.012}$&$ 0.023$&$\mathbf{0.013}$&$\mathbf{0.020}$&$\mathbf{0.036}$\\ 
\hline
\hline
\parbox[t]{2mm}{\multirow{6}{*}{\rotatebox[origin=c]{90}{ $N=2^9$}}}&\parbox[t]{1mm}{\multirow{2}{*}{\rotatebox[origin=c]{90}{m}}}& LF&$ \mathbf{-0.020}$&$\mathbf{-0.040}$&$\mathbf{-0.075}$&$-0.037$&$-0.063$&$  \mathbf{-0.100}$\\ 
 & & MMSE&$-0.019$&$-0.036$&$ -0.070$&$\mathbf{-0.031}$&$\mathbf{-0.060}$&$ -0.120$\\ 
\cline{2-9}
& \parbox[t]{1mm}{\multirow{2}{*}{\rotatebox[origin=c]{90}{s}}}& LF&$0.006$&$0.009$&$ 0.014$&$0.010$&$ 0.014$&$  0.020$\\ 
 & & MMSE&$\mathbf{0.004}$&$\mathbf{0.006}$&$  \mathbf{0.010}$&$\mathbf{0.004}$&$\mathbf{0.007}$&$\mathbf{0.013}$\\ 
\cline{2-9}
& \parbox[t]{1mm}{\multirow{2}{*}{\rotatebox[origin=c]{90}{rms}}}& LF&$0.006$&$0.009$&$ 0.015$&$  0.020$&$ 0.027$&$\mathbf{0.032}$\\ 
 & & MMSE&$\mathbf{0.004}$&$\mathbf{0.007}$&$ 0.015$&$\mathbf{0.012}$&$\mathbf{0.021}$&$ 0.038$\\ 
\hline
\end{tabular}
\end{table}

\noindent\textbf{Compound Poisson Cascade (CPC).} \quad CPCs were introduced to overcome certain limitations of the CMCs that are caused by their discrete split-and-multiply construction \cite{MultiProducts,IDCChainais}. 
In the construction of CPCs, the localization of the multipliers in the space-scale volume follows a Poisson random process with specific prescribed density. We use CPCs with log-normal multipliers $W=\exp(Y)$, where $Y\sim\mathcal{N}(\mu,\sigma)$
is a Gaussian random variable (CPC-LN), or log-Poisson CPCs for which multipliers $W$ are reduced to a constant $w$ (CPC-LP).
The first log-cumulants of CPC-LN are given by
$c_1=-\left(\mu+1-\exp\left(\mu +\sigma^2/2\right)\right) { +  \alpha}$, 
$c_2=-(\mu^2+\sigma^2)$, and $\,c_p\neq0$ for  $p\geq3$.
Here, we fix $\mu=-0.1$. For CPC-LP,  $c_2=-\myln(w)^2$.

\noindent\textbf{Fractional Brownian motion (fBm).}\quad We use 2D fBms as defined in \cite{Stein02}. FBm is not a CMC process and is based on an additive construction instead. Its multifractal and statistical properties are entirely determined by a single parameter $H$ such that $c_1=H$, $c_2= 0$ and $c_p=0$ for all $p> 2$,  and below we set $H=0.7$. 

\subsection{Numerical simulations}
\noindent\textbf{Wavelet transform.}\quad A Daubechies's mother wavelet with $N_{\psi}=2$ vanishing moments is used, and $\alpha=1$ in \eqref{equ:fi}, which is sufficient to ensure positive uniform regularity for all processes considered.

\noindent\textbf{Estimation.}\quad 
The linear regression weights $w_j$  in the standard estimator \eqref{equ:c2LF} have to satisfy the usual constraints $ \sum_{j_1}^{j_2} j w_{j}=  1$ and  $\sum_{j_1}^{j_2} w_{j}  =  0$ and can be chosen to reflect the confidence granted to each $\widehat{\mbox{Var}}_{n_j}[\myln \Lead(j,\cdot)]$, see \cite{Wendt2007d,Wendt20091100}. 
Here, they are chosen proportional to $n_j$ as suggested in \cite{Wendt2007d}.
The linear regression based standard estimator \eqref{equ:c2LF} will be denoted LF (for ``linear fit'') in what follows.
The Gibbs sampler is run with $N_{\textrm{mc}}=7000$ iterations and a burn-in period of $N_{\textrm{bi}}=3000$ samples.
The  bandwidth parameter $\eta$ in \eqref{fullWhittleMod} has been set to $\eta=0.3$ following preliminary numerical simulations; these are illustrated in Fig. \ref{fig:bandwidth} where estimation performance is plotted as a function of $\eta$ for LN-CMC ($N=2^8$ top, $N=2^9$ bottom) with two different values of $c_2$ ($-0.02$ in red, $-0.08$ in blue). As expected, $\eta$ tunes a classical bias-variance tradeoff: a large value of $\eta$ leads to a large bias and small standard deviation and vice versa. The choice $\eta=0.3$ yields a robust practical compromise.

\noindent\textbf{Performance assessment.}\quad 
We apply the LF estimator \eqref{equ:c2LF} and the proposed MAP and MMSE estimators \eqref{equ:map} and \eqref{equ:mmse} to $R=100$ independent realizations of size $N\times N$ each for the above described multifractal processes. 
A range of weak to strong multifractality parameter values $c_2\in\{-0.01, -0.02, -0.04, -0.06, -0.08\}$ and sample sizes  $N\in\{2^6,2^7,2^8,2^9\}$ are used.
The coarsest scale $j_2$ used for estimation is set such that $n_{j_2}\ge100$ (i.e., the coarsest available scale is discarded), yielding $j_2=\{2,3,4,5\}$, respectively, for the considered sample sizes. The finest scale $j_1$ is commonly set to $j_1=2$ in order to avoid pollution from improper initialization of the wavelet transform, see \cite{Veitch2000} for details. 
Performance is evaluated using the sample mean, the sample standard deviation and the root mean squared error (RMSE) of the estimates averaged across realizations
\begin{equation*}
\textrm{m}=\widehat{\EE}[\hat c_2],\;
\textrm{s}=\sqrt{\widehat{\mbox{Var}}[\hat c_2]}, \;
\textrm{rms}=\sqrt{(m-c_2)^2+s^2}.
\end{equation*}

\begin{table}[!t]
\setlength{\tabcolsep}{0.7mm}
\renewcommand{\arraystretch}{0.95}
\caption{\label{tab:SMALL}Estimation performance for CMC-LN ($\textnormal{a}$) and CPC-LN ($\textnormal{b}$) for sample sizes $N=\{2^6,2^7\}$ and $j_1=1$, $j_2=\{2,3\}$. Best results are marked in bold.}
\centering
\begin{tabular}{c|c|c||r|r|r|r|r}
\hline
 \multicolumn{8}{l}{(a)\hskip3.2cm CMC-LN}\\
\hline
\multicolumn{3}{c||}{$c_2$}&$ -0.01\;\;$&$ -0.02\;\;$&$ -0.04\;\;$&$ -0.06\;\;$&$ -0.08\;\;$\\ 
        \hline
    \parbox[t]{2mm}{\multirow{6}{*}{\rotatebox[origin=c]{90}{ $N=2^6$}}}&\parbox[t]{2mm}{\multirow{2}{*}{\rotatebox[origin=c]{90}{m}}}& LF &$-0.042$&$-0.051$&$-0.067$&$-0.082$&$ -0.110$\\
   & & MMSE &$\mathbf{-0.014}$&$\mathbf{-0.022}$&$\mathbf{-0.038}$&$\mathbf{-0.059}$&$\mathbf{-0.078}$\\
        \cline{2-8}
    &\parbox[t]{2mm}{\multirow{2}{*}{\rotatebox[origin=c]{90}{s}}}& LF &$ 0.024$&$ 0.030$&$ 0.042$&$ 0.042$&$ 0.070$\\
    & & MMSE &$ \mathbf{0.010}$&$\mathbf{0.014}$&$\mathbf{0.018}$&$\mathbf{0.026}$&$\mathbf{0.038}$\\
         \cline{2-8}
    &\parbox[t]{2mm}{\multirow{2}{*}{\rotatebox[origin=c]{90}{rms}}}& LF &$ 0.040$&$ 0.043$&$ 0.050$&$ 0.047$&$ 0.076$\\
    & & MMSE &$\mathbf{0.010}$&$\mathbf{0.014}$&$\mathbf{0.018}$&$\mathbf{0.026}$&$\mathbf{0.038}$\\
    \hline \hline
       \parbox[t]{2mm}{\multirow{6}{*}{\rotatebox[origin=c]{90}{ $N=2^7$}}}&\parbox[t]{1mm}{\multirow{2}{*}{\rotatebox[origin=c]{90}{m}}}& LF &$-0.035$&$-0.044$&$-0.064$&$-0.082$&$  -0.100$\\
   & & MMSE &$\mathbf{-0.013}$&$\mathbf{-0.023}$&$\mathbf{-0.044}$&$\mathbf{-0.064}$&$\mathbf{-0.082}$\\
        \cline{2-8}
    &\parbox[t]{1mm}{\multirow{2}{*}{\rotatebox[origin=c]{90}{s}}}& LF &$ 0.010$&$ 0.013$&$ 0.019$&$ 0.024$&$ 0.026$\\
    & & MMSE &$\mathbf{0.005}$&$\mathbf{0.008}$&$\mathbf{0.013}$&$\mathbf{0.017}$&$ \mathbf{0.018}$\\
         \cline{2-8}
    &\parbox[t]{1mm}{\multirow{2}{*}{\rotatebox[origin=c]{90}{rms}}}& LF &$ 0.027$&$ 0.027$&$ 0.031$&$ 0.033$&$ 0.033$\\
    & & MMSE &$\mathbf{0.006}$&$\mathbf{0.009}$&$\mathbf{0.014}$&$\mathbf{0.017}$&$ \mathbf{0.018}$\\
    \hline
\hline
 \multicolumn{8}{l}{(b)\hskip3.2cm CPC-LN}\\
\hline
\multicolumn{3}{c||}{$c_2$}&$ -0.01\;\;$&$ -0.02\;\;$&$ -0.04\;\;$&$ -0.06\;\;$&$ -0.08\;\;$\\ 
\hline
\parbox[t]{2mm}{\multirow{6}{*}{\rotatebox[origin=c]{90}{ $N=2^6$}}}&\parbox[t]{1mm}{\multirow{2}{*}{\rotatebox[origin=c]{90}{m}}}& LF&$-0.026$&$-0.054$&$-0.076$&$-0.085$&$  -0.100$\\ 
 & & MMSE&$\mathbf{-0.0082}$&$\mathbf{-0.017}$&$\mathbf{-0.030}$&$ \mathbf{-0.050}$&$\mathbf{-0.065}$\\ 
\cline{2-8}
& \parbox[t]{1mm}{\multirow{2}{*}{\rotatebox[origin=c]{90}{s}}}& LF&$ 0.024$&$ 0.029$&$ 0.045$&$ 0.068$&$ 0.067$\\ 
 & & MMSE&$\mathbf{0.005}$&$\mathbf{0.011}$&$\mathbf{0.018}$&$ \mathbf{0.028}$&$\mathbf{0.033}$\\ 
\cline{2-8}
& \parbox[t]{1mm}{\multirow{2}{*}{\rotatebox[origin=c]{90}{rms}}}& LF&$ 0.029$&$ 0.044$&$ 0.058$&$ 0.073$&$  0.070$\\ 
 & & MMSE&$\mathbf{0.006}$&$\mathbf{0.011}$&$\mathbf{0.021}$&$  \mathbf{0.030}$&$\mathbf{0.036}$\\ 
\hline
\hline
\parbox[t]{2mm}{\multirow{6}{*}{\rotatebox[origin=c]{90}{ $N=2^7$}}}&\parbox[t]{1mm}{\multirow{2}{*}{\rotatebox[origin=c]{90}{m}}}& LF&$-0.021$&$-0.047$&$-0.064$&$-0.082$&$ -0.110$\\ 
 & & MMSE&$\mathbf{-0.008}$&$\mathbf{-0.017}$&$\mathbf{-0.035}$&$\mathbf{-0.057}$&$\mathbf{-0.079}$\\ 
\cline{2-8}
& \parbox[t]{1mm}{\multirow{2}{*}{\rotatebox[origin=c]{90}{s}}}& LF&$0.0091$&$ 0.013$&$  0.020$&$ 0.024$&$ 0.032$\\ 
 & & MMSE&$\mathbf{0.0032}$&$\mathbf{0.0082}$&$\mathbf{0.012}$&$\mathbf{ 0.018}$&$\mathbf{0.021}$\\ 
\cline{2-8}
& \parbox[t]{1mm}{\multirow{2}{*}{\rotatebox[origin=c]{90}{rms}}}& LF&$ 0.014$&$  0.030$&$ 0.031$&$ 0.033$&$ 0.042$\\ 
 & & MMSE&$\mathbf{0.004}$&$\mathbf{0.0087}$&$\mathbf{0.013}$&$ \mathbf{0.019}$&$\mathbf{0.021}$\\ 
\hline
\end{tabular}
\end{table}

\begin{table}
\renewcommand{\arraystretch}{0.95}
\caption{FBm estimation performance for sample sizes $N=\{2^7,2^8,2^9\}$ and $j_1=2$, $j_2=\{3,4,5\}$. Best results are marked in bold.}
\centering
\begin{tabular}{c|c||r|r|r}
    \hline
\multicolumn{2}{c||}{$N$}&$2^7$&$2^8$&$2^9$\\
\hline    
    \parbox[t]{2mm}{\multirow{2}{*}{\rotatebox[origin=c]{90}{m}}}& LF &$0.0034$&$0.0047$&$0.0037$\\
    & MMSE &$\mathbf{-0.0020}$&$\mathbf{-0.0008}$&$\mathbf{-0.0003}$\\
        \hline
            \parbox[t]{2mm}{\multirow{2}{*}{\rotatebox[origin=c]{90}{s}}}& LF &$ 0.0170$&$0.0089$&$0.0056$\\
    & MMSE &$\mathbf{0.0093}$&$\mathbf{0.0010}$&$\mathbf{0.0002}$\\
        \hline
            \parbox[t]{2mm}{\multirow{2}{*}{\rotatebox[origin=c]{90}{rms}}}& LF &$ 0.0180$&$  0.0100$&$0.0067$\\
    & MMSE &$\mathbf{0.0095}$&$\mathbf{0.0012}$&$\mathbf{0.0004}$\\
        \hline
\end{tabular}
\label{tab:FBM}
\end{table}

\begin{figure*}[t]
\centering
\includegraphics[width=1\linewidth]{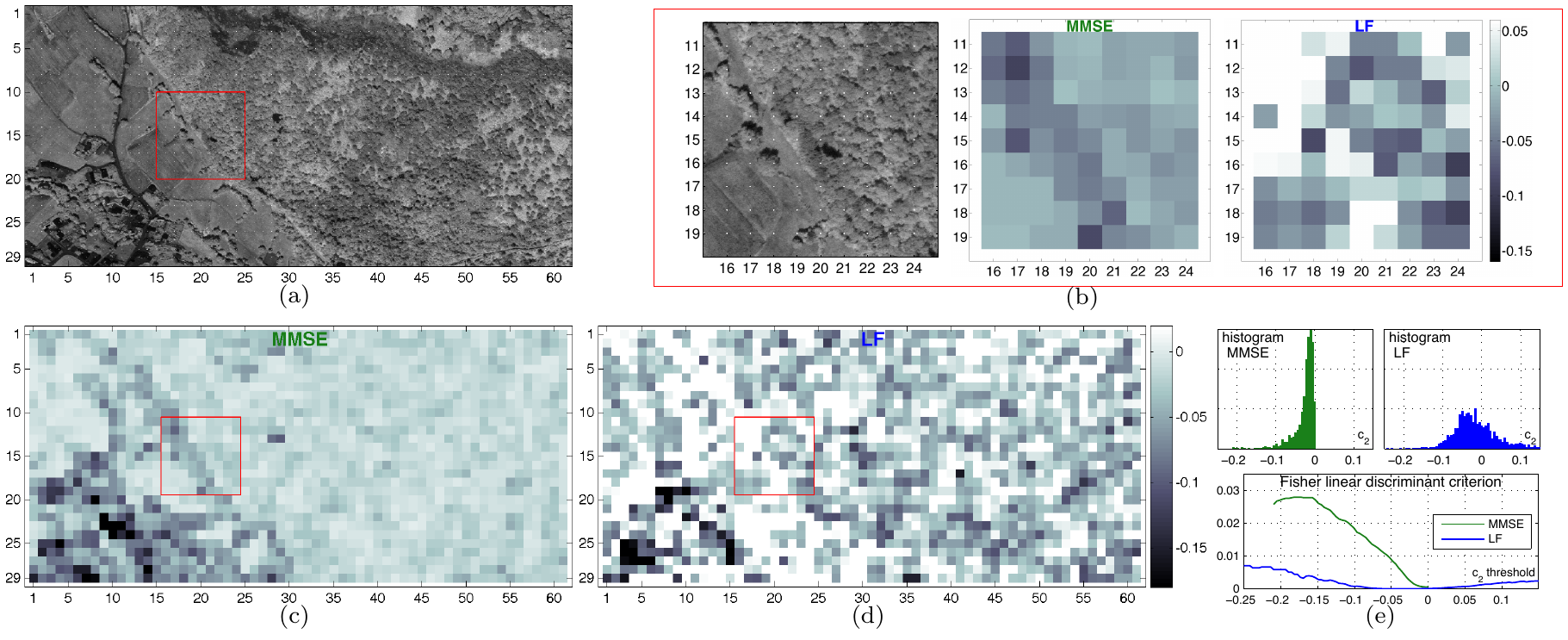}\vspace{-2mm}
\caption{\label{fig:HS} 
Band \#20 of a hyperspectral datacube (a); estimates of $c_2$ for overlapping $64\times64$ pixel patches obtained by MMSE (c) and LF (d); zooms on the patches indicated by a red frame (b); the centers of the image patches are indicated by white dots in the original image, the distance between two of the dots corresponds to one half of the patch size, axis labels indicate patch numbers. Histograms and Fisher linear discriminant criteria for estimates of $c_2$ obtained by MMSE and LF (e).
}
\end{figure*}

\subsection{Results}

\noindent\textbf{Estimation performance.}\quad
Tab. \ref{tab:LARGE} summarizes the estimation performance of LF and MMSE estimators for CMC-LN, CPC-LN, CMC-LP, CPC-LP (subtables (a)-(d), respectively) and for sample sizes $N=\{2^8,2^9\}$. 
The performance of the MAP estimator was found to be similar to the MMSE estimator and therefore is not reproduced here due to space constraints. 
Note, however, that different (application dependent) priors for $\gammab$ may lead to different results (here, the non-informative prior \eqref{equ:prior} is used). 

First, it is observed that the proposed algorithm slightly but systematically outperforms LF in terms of bias. This reduction of bias does not depend on a specific choice of the multifractal process or its parameters, or on the sample size.
Second, and most strikingly, the proposed Bayesian estimators yield significantly reduced standard deviations, with a reduction of up to a factor of $3$ as compared to linear regressions. The standard deviation reduction is more important for small values of $|c_2|$ yet remains above a factor of $1.5$ for large values of $|c_2|$.

These performance gains are directly reflected in the overall RMSE values, which remain up to a factor of $2.5$ below those of linear fits.
Finally, note that the estimation performance for CMCs and CPCs with log-Poisson multipliers are found to be slightly inferior to those with log-normal multipliers. This may be due to an arguably slightly stronger departure from Gaussian for the former, cf. Fig. \ref{fig:cov}. 

\noindent\textbf{Performance for small sample size.}\quad
For small sample sizes $N\leq 2^7$, the limited number of available scales forces the choice $j_1=1$.
Results for $N=\{2^6,2^7\}$ (for which $j_2=\{2,3\}$, respectively) are reported in Tab. \ref{tab:SMALL}. They indicate that the performance gains of the proposed Bayesian estimators with respect to LF estimators are even more pronounced for small sample size, both in terms of bias and standard deviations, yielding a reduction of RMSE values of up to a factor of $4$. 
In particular, note that LF yields biases that are prohibitively large to be useful in real-world applications due to the use of the finest scale $j=1$, cf.,  \cite{Wendt20091100}. 
Notably, values $c_2=0$ cannot be reliably detected with LF.
In contrast, the proposed Bayesian procedure yields sufficiently small bias and standard deviations to enable the estimation of the multifractality parameter $c_2$ even for very small images (or image patches) of size $64\times64$.
The reported performance gains come at the price of an increased computational cost, with computation times of the order of $8$s ($N=64$) to $50$s ($N=512$) per image, respectively, on a standard desktop computer,
which is two orders of magnitude larger than the computational cost of the LF estimator.
 
\noindent\textbf{Performance for fractional Brownian motion.}\quad
Self-similar fBms with $c_2=0$ do not belong to the class of MMC processes for which the proposed estimation procedure was designed. The correlation structure of the wavelet coefficients of fBms has been studied in, e.g., \cite{Flandrin1992}. This correlation is weak, i.e., it goes to zero fast with the distance between wavelet coefficients in the time-scale plane.
FBm results are summarized in Tab. \ref{tab:FBM}. They indicate that the performance of the LF estimator is comparable to the case $c_2=-0.01$ reported in Tab. \ref{tab:LARGE}. 
In contrast, the proposed Bayesian estimators are practically unbiased and have standard deviations and RMSE values that significantly outperform those of LF by up to a factor $10$. 
Therefore, it is much more likely to be able to identify a model for which $c_2 = 0$ when using the proposed Bayesian procedure instead of the classical LF.

\section{Illustration for real-world data}
\label{sec:Madonna}
We illustrate the proposed Bayesian estimation procedure for the multifractal analysis of a real-world image in Fig. \ref{fig:HS}(a).
The image of size $960 \times 1952$ pixels is the channel \#20 of a  hyperspectral datacube corresponding to a forested area near a city that was acquired by the Hyspex hyperspectral scanner over Villelongue, France, during the Madonna project \cite{MADONNA}. Estimates of $c_2$ are computed for $29\times60$ overlapping patches of size $64\times64$ pixels.

The estimates are plotted in Fig. \ref{fig:HS} for MMSE (c) and LF (d), subfigure (b) provides a magnification (indicated by a red frame) on the square of patches of rows 11-19 / columns 16-24.
%
Visual inspection indicates that the Bayesian estimates are much better reproducing the spatial structure of the image texture than the classical LF (cf., Fig. \ref{fig:HS}(a), (c) and (d)). 
Specifically, the zoom in Fig. \ref{fig:HS}(b) (equivalently, the corresponding textures in Fig. \ref{fig:HS}(a) and (c)),
shows that the Bayesian estimates are spatially strongly homogeneous for the forested regions with visually homogeneous texture (e.g., upper right portions in Fig. \ref{fig:HS}(b)), indicating a weak yet non-zero multifractality for these regions. Similar observations are obtained for other homogeneous vegetation patches (e.g., bottom left corners in Fig. \ref{fig:HS}(b)). 
Moreover, the zones of mixed vegetation (e.g., upper left corner in Fig. \ref{fig:HS}(b)) also yield spatially coherent and consistent estimates of $c_2$, with more negative values (stronger multifractality).
The LF based estimates display a strong variability throughout the image. Indeed, even for the homogeneous texture in the forested regions, LF yields strongly spatially varying estimates.
Finally, note that the strongly negative values of $c_2$ observed for both MMSE and LF in the bottom left corner of Fig. \ref{fig:HS}(c) correspond to regions consisting of both (textured) vegetations and of roofs of buildings (with close to zero amplitudes and no texture).

Although no ground truth is available for this illustration, a more quantitative analysis of the relative quality of estimates of $c_2$ obtained with MMSE and LF is proposed here.
First, the reference-free image quality indicator of \cite{blanchet2012explicit}, which quantifies the image sharpness by approximating a contrast-invariant measure of its phase coherence \cite{Blanchet2008icip}, is calculated for the maps of $c_2$ in Fig. \ref{fig:HS}(c) and (d). These sharpness indexes are $10.8$ for MMSE and a considerably smaller value of $4.6$ for LF, hence reinforcing the visual inspection-based conclusions of improved spatial coherence for MMSE described above.
Second, Fig. \ref{fig:HS}(e) (top) shows histograms of the estimates of $c_2$ obtained with MMSE and LF, which confirm the above conclusions of significantly larger variability (variance) of LF as compared to MMSE. 
Moreover, LF yields a large portion of estimates with positive values, which are not coherent with multifractal theory since necessarily $c_2<0$, while MMSE estimates are consistently negative. 
Finally, the Fisher linear discriminant criterion \cite[Ch. 3.8]{duda2012pattern} is calculated for $c_2$ obtained with MMSE and with LF, as a function of a threshold for $c_2$ separating two classes of textures. The results, plotted in Fig. \ref{fig:HS}(e) (bottom), indicate that the estimates obtained with MMSE have a far superior discriminative power than those obtained with LF.


\section{Conclusions}
\label{sec:conclusions}

This paper proposed a Bayesian estimation procedure for the multifractality parameter of images. The procedure relies on the use of novel multiresolution quantities that have recently been introduced for regularity characterization and multifractal analysis, i.e., wavelet leaders. A Bayesian inference scheme was enabled through the formulation of an empirical yet generic semi-parametric statistical model for the logarithm of wavelet leaders. This model accounts for the constraints imposed by multifractal theory and is designed for a large class of multifractal model processes.
The Bayesian estimators associated with the posterior distribution of this model were approximated by means of samples generated by a Metropolis-within-Gibbs sampling procedure, wherein the practically infeasible evaluation of the exact likelihood was replaced by a suitable Whittle approximation. 
The proposed procedure constitutes, to the best of our knowledge, the first operational Bayesian estimator for the multifractality parameter that is applicable to real-world images and effective both for small and large sample sizes.
Its performance was assessed numerically using a large number of multifractal processes for several sample sizes. The procedure yields improvements in RMSE of up to a factor of $4$ for multifractal processes, and up to a factor of $10$ for fBms when compared to the current benchmark estimator. 
The procedure therefore enables, for the first time, the reliable estimation of the multifractality parameter for images or image patches of size equal to $64\times64$ pixels.
It is interesting to note that the Bayesian framework introduced in this paper could be generalized to hierarchical models, for instance, using spatial regularization for patch-wise estimates.
In a similar vein, future work will include the study of appropriate models for the analysis of multivariate data, notably for hyperspectral imaging applications. 

\bibliographystyle{plain}
\bibliography{bibi_cbx}

\end{document}